\input phyzzx
\overfullrule=0pt
\tolerance=5000
\twelvepoint

\date{hep-th/0005074}
\titlepage
\title{The Ramond-Ramond self-dual Five-form's Partition Function on $T^{10}$}
\vglue-.25in
\author{ Louise Dolan \foot{Research supported in part by the U.S. Department
of Energy under Grant No. DE-FG 05-85ER40219/Task A. 
Email: dolan@physics.unc.edu}} 
\address{Department of Physics and Astronomy
\break University of North Carolina
\break Chapel Hill
\break North Carolina 27599-3255}
\author{Chiara R. Nappi\foot{Research supported in part by
DOE grant DE-FG03-84ER40168. Email: nappi@usc.edu}}
\address{Department of Physics and Astronomy
\break and Caltech-USC Center for Theoretical Physics
\break University of Southern California
\break Los Angeles, CA 90089-2535}
\medskip

\abstract{In view of the recent interest in formulating a quantum
theory of Ramond-Ramond p-forms, we exhibit an $SL(10,{\cal Z})$ 
invariant partition function for the  chiral four-form
of Type IIB string theory on the ten-torus. We follow the 
strategy used to derive a modular invariant partition function
for the chiral two-form of the M-theory fivebrane. 
We also generalize the calculation to self-dual quantum fields 
in  spacetime dimension $2p=2+4k$, and 
display the $SL(2p,{\cal Z})$ automorphic forms for odd $p>1$.
We relate our explicit calculation  
to a computation of the B-cycle periods, which are  
discussed in the work of Witten.} 

\vfill\eject

\REF\DN{L. Dolan and C. R. Nappi, ``A Modular Invariant Partition Function for 
the Fivebrane'', Nucl. Phys. {\bf B530} (1998) 683; hep-th/9806016.} 
\REF\Wittentwo{ E. Witten, ``Duality Relations Among Topological Effects in
String Theory'', hep-th/9912086.}
\REF\WM{ G. Moore and E. Witten, ``Self-Duality, Ramond-Ramond Fields,
and K-Theory, hep-th/0012279.}
\REF\Freed{ D. Freed and M. Hopkins, ``On Ramond-Ramond Fields and K-Theory'',
hep-th/0002027.}
\REF\Witten{E. Witten, ``Five-brane Effective Action in M-theory,''
J. Geom. Phys. {\bf 22} (1997) 103; hep-th/9610234.}
\REF\hop {M. Hopkins and I. M. Singer, ``Quadratic Functions in Geometry,
Topology, and M-Theory'', to appear.} 
\REF\Kervaire{ M. Kervaire, ``A Manifold That Does Not Admit Any
Differentiable Structure'', Comment. Math. Helv. {\bf 34} (1960) 257-270.}   
\REF\Green{M.B. Green, J. H. Schwarz and E. Witten,
{\it Superstring Theory},
vol. I and II, Cambridge University Press: Cambridge, U.K. 1987.} 
\REF\henning{M. Henningson, B. Nilsson and P. Salomonson,
``Holomorphic Factorization of Correlation Functions in (4k+2)-dimensional
(2k)-form Gauge Theory'', JHEP {\bf 9909:008} (1999); hep-th/9908107.}
\REF\Schwarz{ J.H. Schwarz. ``Coupling a Self-dual Tensor to
Gravity in Six Dimensions,'' Phys. Lett. {\bf 395} (1997) 191; 
hep-th/ 9701008.}
\REF\Mumford {D. Mumford, {\it Tata Lectures on Theta} vols. I and II,
Boston: Birkhauser 1983; L. Alvarez-Gaume, G. Moore, C. Vafa,
``Theta Functions, Modular Invariance, and Strings'', Comm. Math. Phys.
{\bf 106} (1986) 1-40;  P. Sarnak, {\it Some Applications of Modular Forms},
Cambridge, 1990.}  
\REF\Coxeter{ H.S.M. Coxeter and W.O.J. Moser,
{\it Generators and Relations for Discrete Groups},
Springer-Verlag: New York 1980.}

\chapter {Introduction}

The calculation reported in this paper extends the methods in [\DN].
The result is of interest as it involves a Ramond-Ramond chiral form
(while in [\DN] we dealt with a chiral-two form in six dimensions which
is a NS form). 
Although restricted to the ten-torus, the result is useful in view of  
the analysis in [\Wittentwo, \WM, \Freed] which 
sets out a method for calculating the partition functions of 
self-dual Ramond-Ramond fields in Type II string theory on  
general manifolds and emphasizes their central role
in formulating a quantum theory of non-Lagrangian fields.

Our calculation also provides an explicit formula for the 
partition function of the self-dual p-form $G_p$ on $T^{2p}$
when the dimension of spacetime $2p = 2+4k$ is twice odd,
(for odd $p>1$). These dimensions occur since the Hodge 
star operator $\ast$ can be used to define a self-dual $p=(1+2k)$-form
for Minkowski signature metrics\foot{If we define the
Hodge dual operation which converts a d-form into a (D-d)-form
as $( ^*J)_{M_1\ldots M_{D-d}} \equiv
{1\over d!} \epsilon_{M_1\ldots M_{D-d}\mu_1\ldots\mu_d} J^{\mu_1\ldots\mu_d}
\, {1\over {\sqrt {| g_D |}}}$
where
$\epsilon_{M_1\ldots M_{D-d}\mu_1\ldots\mu_d}
\equiv g_{M_1N_1}\ldots g_{M_{D-d}N_{D-d}} g_{\mu_1\nu_1} g_{\mu_d\nu_d}
\epsilon^{N_1\dots N_{D-d}\nu_1\ldots \nu_d}$,
then $( ^{**}J)^{\mu_1\ldots \mu_d} = 
(-1)^{(D-d)d}\, g_D\, {1\over {| g_D |}}\, J^{\mu_1\ldots\mu_d}$. So for
Minkowski signature metrics, which have $|g_D|= - g_D$, we have $D = 2d$ 
with odd $d$ for $^{**} =1$.}  
(which are needed for quantum mechanics) and has $\ast \ast =1$.  
The dimensions $(2+4k)$ are the same as those for which 
Kervaire invariants are defined. The question of modular invariance
of the partition function of self-dual quantum fields on general 
$(2+4k)$-dimensional spacetimes can be related to these invariants
through the work of [\Witten, \hop, \Kervaire].

For the torus compactifications, we identify the actual value of  
the determinant $\Delta$ describing the contribution of the non-zero modes,
where the partition function is $Z = {\Theta\over\Delta}$. From
the sum over zero modes, we find the theta function $\Theta$
and an explicit representation of the period matrix,
which alternatively could be derived by a 
direct assignment of the B-cycle periods [\Wittentwo].
The ten torus example corresponds to zero shift in the Dirac quantization
condition for the five-form field strength $G_5$, 
so it can be formulated without applying the elegant K-theory techniques of 
[\Wittentwo, \WM]. Thus our examples do not probe certain fundamental 
properties 
of self-dual RR fields on arbitrary manifolds, but nonetheless provide 
an explicit case of B-lattice basis vectors, which are somewhat involved 
to calculate directly even on the flat torus with a general metric. 
Given the complicated nature of actual constructions on general manifolds, 
we display the $SL(2+4k,{\cal Z})$ automorphic forms
for quantum self-dual fields on the torus,
although the relevance
to string theory for $k>2$ is not apparent.

The first step is, of course, to define the partition function. Given the 
notorious difficulties in writing down covariant Lagrangians for a chiral form, 
we will not start with a Lagrangian (which we do not have) 
and write down the path integral. 
Rather, as usual in string calculations [\Green],
we compute the trace ${\rm tr} e^{-t{\cal H}}$ on the
twisted ten-torus. 
To that purpose we define the Hamiltonian and the momenta
for our theory along the lines of [\DN]. Our formalism for the
Ramond-Ramond self-dual five-form partition function will be manifestly
$SL(9,\cal{Z})$ invariant, 
which combined with an extra $SL(2,{\cal Z})$ symmetry we show, 
will ensure its  $SL(10,{\cal Z})$ modular invariance on the ten-torus. 

While the partition function for the 
two-dimensional chiral boson in two dimensions 
is not modular invariant, in [1] we proved that the chiral two-form of the 
M-theory fivebrane is $SL(6,{\cal Z)}$ modular invariant on the six-torus.
That proof is generalizable to any chiral
$2k$-form, for integer $k>0$, which corresponds to a $(2k+1)$-form 
self-dual field strength in $2+4k$ dimensions. 
A crucial step is to view  $T^{2+4k}$  as the product $T^2 \times
T^{4k}$. The extra $SL(2,{\cal Z})$ symmetry mentioned above is the one 
associated with the two-torus $T^2$.

As usual, the calculation of the trace factorizes as 
$Z= Z_{\rm osc} \cdot Z_{\rm zero\, modes}$, where
$Z_{\rm osc}$ is the contribution from the sum over the oscillators
and  $Z_{\rm zero\, modes}$ comes from the sum over the zero modes.
The sum over the zero modes will have an anomaly under
$SL(2,{\cal Z})$ which cancels against a similar anomaly in the  
sum over oscillators.

The sum over the oscillators associated with the three degrees of freedom 
of the two-form potential of the M-theory fivebrane
factored into sums over ``massless'' and ``massive'' modes,
according to whether they had zero momentum or non-zero momentum on $T^4$. 
The $SL(2,{\cal Z})$ 
anomaly needed to cancel the anomaly coming from the zero modes was  due 
to the contribution of the `2d-massless' modes, {\it i.e} the modes
with zero momentum in the transverse directions on $T^4$. 
The modes with non-zero momentum 
on $T^4$   behaved like  massive scalars in two dimensions, 
and did not have any anomaly.
The overall effect was that the three degrees of freedom of the chiral
two-form mimicked the situation of three non-chiral bosons. 
A similar thing will happen  with the chiral $2k$-form studied in this paper. 
The compactificaton on $T^{10}$ can be viewed as compactification on
$T^2\times T^8$. Effectively,
the thirty-five degrees of freedom of the four-form potential (which is the
${\bf{35_c}}$ representation of the Spin(8) 
little group for 10d massless states)
behave like thirty-five 2d non-chiral scalars.
 
Our formula has the property that in the case of $T^{2p}$ for odd $p>1$, 
there is no spin structure dependence.
The situation can be different for compactifications
on more general manifolds [\Witten, \hop, \henning]. 
A general analysis of the correlation functions of $2k$-chiral forms, 
including the $SL(2+4k,{\cal Z})$ modular invariance 
of the torus compactification,
is provided by Henningson, Nilsson, and Salomonson in [\henning] using
holomorphic factorization.

Section 2 describes the application of the method of [\DN] to
the Ramond-Ramond chiral four-form of Type IIB string theory. 
Section 3 generalizes our results to spacetime dimension $2+4k$.
In sections 3 and 4, from our calculation of the partition function,
we identify the relevant period matrix,
omega function, and the A- and B-lattices, as 
introduced in [\Wittentwo].
For the case $p=3$, we 
show how for a rectangular torus our period matrix corresponds 
to a direct computation of the periods of the A- and B-cycles,
and this can be extended easily for general $p$. 

\chapter {The $G_5$ Partition Function on $T^{10}$}
We start with a chiral four-form in ten dimensions 
$B_{MNRS}$ with a self-dual five-form field strength 
$H_{LMNRS}= \partial_LB_{MNRS} + \partial_MB_{NRSL} + \partial_NB_{RSLM}
+\partial_RB_{SLMN} +\partial_SB_{LMNR}$ 
($1\le L,M,N \le 10$),
which satisfies $$H_{LMNRS}(\vec\theta,\theta^{10})= 
{1\over{{5!}\sqrt{-G}}}G_{LL'}G_{MM'}G_{NN'}
G_{RR'}G_{SS'}
\epsilon^{L'M'N'R'S'PQTUV}H_{PQTUV}(\vec\theta,\theta^{10}) \eqn\A $$
where $G_{MN}$ is the 10d metric.
Following [\Schwarz, \DN], we use this equation of motion 
to eliminate the components $H_{10\, mnrs}$
in terms of the other components $H_{lmnrs}$ with $l,m,n,r,s
=1...9$.
The independent fields $H_{lmnpq}$
form a totally antisymmetric tensor which is no longer self-dual.
To describe the partition function of the chiral four-form,
we use their 
Hamiltonian and momenta which are written in a 9d covariant way:
$${\cal H} = {1\over 2 (5!)}{\int_0}^{2\pi}
d\theta^1...d\theta^9{\sqrt G_9}{G_9}^{ll'}
{G_9}^{mm'}{G_9}^{nn'} {G_9}^{pp'} {G_9}^{qq'} 
H_{lmnpq}(\vec\theta,\theta^{10})\,
H_{l'm'n'p'q'}(\vec\theta,\theta^{10}) \eqn\Ham$$
$$P_l = -{5\over 2 (5!)^2}{\int_0}^{2\pi}
d\theta^1...d\theta^9 \epsilon^{rsvwumnpq} H_{umnpq}(\vec\theta,\theta^{10})\,  
H_{lrsvw}(\vec\theta,\theta^{10})\eqn\Pel$$  
where now indices are raised with the 9d-metric $G_9^{mn}$ and 
$\epsilon_{123456789} \equiv G_9\epsilon^{123456789} = G_9$.

An arbitrary 
flat metric $G_{MN}$ on $T^{10}$ is a function of 55 parameters and 
can be represented by the line element
$$\eqalign{ds^2 = & {R_1}^2(d\theta^1 -\alpha d\theta^{10})^2 +
{R_{10}}^2(d\theta^{10})^2 \cr
+&\sum_{i,j=2...9}g_{ij}(d\theta^i -
\beta^id\theta^1-\gamma^id\theta^{10})
(d\theta^j - \beta^jd\theta^1 - \gamma^jd\theta^{10})\,. \cr}\eqn\B$$
Above,  $0\le\theta^I\le 2\pi$, $1\le I\le 10$ and 
we have singled out  directions 1 and 10, in accordance with viewing
$T^{10}$ as $T^2\times T^8$. The  direction 10 will be our time direction.
The 55 parameters are $R_1$, $R_{10}$,
$g_{ij}$ (an 8d metric), $\beta^i,\gamma^j$ (the angles 
between directions  1 and $i,j$ respectively); and  
$\alpha$ is related to the angle between 1 and 10.
As in string theory [\Green],  the partition function 
is given in terms of the Hamiltonian and momenta  by 
$$ Z(R_1, R_{10}, g_{ij}, \alpha,\beta^i, \gamma^i) = 
tr  \exp\{-t{\cal H} + i2\pi\alpha P_1 + i2\pi (\alpha\beta^i +
\gamma^i)P_i\} \eqn\trace$$ where $t=2\pi R_{10}$. 
(Notice the metric in {\B} is one that has been rotated to 
Euclidean signature.) 
The partition function \trace\ is by construction $SL(9,{\cal Z})$
invariant, due to the underlying $SO(9)$ invariance
in the coordinate space we have labelled $l=1...9$. We will show it is  
also $SL(2,{\cal Z})$ invariant in the directions 1 and 10. The combination 
of these two invariances yields the $SL(10,{\cal Z})$ invariance of
the automorphic form given by $Z$ in (2.20), along the lines of [\DN].
\vskip20pt

\leftline {\it The Zero Modes of the Partition Function}

To trace on the zero mode operators in \A,
we express the Hamiltonian \Ham\ and momenta \Pel\
in terms of the metric parameters in \B
$$\eqalign{-t{\cal H }+ i2\pi\alpha P_1 + &i2\pi (\alpha\beta^i +
\gamma^i)P_i
= \bigl ( -{\pi \over 5!}  R_{10}R_1{\sqrt  g}g^{ii'}g^{jj'}g^{kk'}
g^{pp'}g^{qq'}H_{ijkpq}H_{i'j'k'p'q'}\cr 
&-{\pi\over (4!)^2 {|\tau |}^2}{R_{10} \over
R_1}{\sqrt g} \, 4!\,g^{[jj'}g^{kk'}g^{pp'}g^{qq']} H_{ijkpq}H_{i'j'k'p'q'}
\gamma^i\gamma^{i'}\cr 
&-\pi {\cal A}^{jkpqj'k'p'q'} {1\over (4!)^2}[H_{1jkpq} + x_{jkpq}] 
[H_{1j'k'p'q'} + x_{j'k'p'q'}]\bigr ) \cr }\eqn\define$$
{\nobreak where $g^{[jj'}g^{kk'}
g^{pp'}g^{qq']}\equiv{\textstyle{1\over 4!}}(g^{jj'}
g^{kk'}g^{pp'}g^{qq'} -\ldots) $ has  $4!$ terms antisymmetric in $jkpq$,}
$$ x_{jkpq} \equiv \beta^i H_{ijkpq} + {i \over (4!)^2} \gamma^i
{{\cal A}^{-1}}_{jkpqj'k'p'q'}  
\epsilon^{j'k'p'q'g h g'h'} H_{ighg'h'}, 
\eqn\some$$ 
$${\cal A}^{jkpqj'k'p'q'}\equiv {R_{10}\over R_1}{\sqrt g}\,
4!\,g^{[jj'}g^{kk'}g^{pp'}g^{qq']} 
+ i\alpha \epsilon^{jkpqj'k'p'q'},\eqn\bene$$
and $\tau \equiv \alpha + i {R_{10} \over R_1}$  
is the modular parameter of the two-torus.

The trace in the partition function \trace\ 
is over all independent Fock space operators which appear in
the normal mode expansion of the free massless tensor gauge field
$B_{MNRS}$. 
The zero mode eigenvalues of the seventy fields $H_{1jkpq}$
are labeled with the integers  $n_1,...,n_{70}$ in  ${\cal Z}^{70}$, 
while the  zero modes of the fifty six fields
$H_{ijkpq}$ are labeled by the integers $n_{71},...,n_{126}$.
Here $2\le i,j,k,p,q,\le 9$.
We make this subdivision  since the $H_{1jkpq}$ remain non-zero in the
``2-d massless'' limit obtained when the momenta on the eight-torus 
are zero, {\it i.e.} $p_i=0$.
Since only the ``2-d massless'' oscillator modes contribute to the
anomaly, only this subset of seventy zero modes will be needed for the 
cancellation. In general, on a flat torus of any dimension $2+4k$, the
zero modes can be separated into $H_{ijkpq\ldots}$ and
$H_{1jkpq\ldots}$. Since the physical degrees of freedom of the
chiral gauge field $B_{MNPQ\ldots}$ are described by the rank $2k$
complex antisymmetric tensor representation of the little group $SO(4k)$,
the number of physical degrees of freedom appearing in the oscillator
calculation will always be equal to the number of zero modes in the
set $H_{1jkpq\ldots}$, so the anomaly will always cancel on the torus.  

The contribution from the zero modes is then 
$$\eqalign{Z_{\rm zero\, modes}  =&  
\sum_{n_{71},\ldots , n_{126}\in {\cal Z}^{56}}
\exp\{-\pi { R_{10}R_1 \over 5!}{\sqrt  g}g^{ii'}g^{jj'}g^{kk'}g^{pp'}g^{qq'}
H_{ijkpq}H_{i'j'k'p'q'}\cr & 
\hskip90pt -{\pi\over (4!)^2 {|\tau |}^2}{R_{10} \over
R_1}{\sqrt g} \,4!\,g^{[jj'}g^{kk'}g^{pp'}g^{qq']} H_{ijkpq}H_{i'j'k'p'q'}
\gamma^i \gamma^{i'}\}\cr & \hskip30pt
\cdot\sum_{n_1,\ldots , n_{70}\in {\cal Z}^{70}} 
e^{-\pi (n+x)\cdot A\cdot (n+x)} \cr}
\eqn\zeroone $$ 
where $A$ is a rank $70$ matrix with 
$A_{11} = {\cal A}^{2345 2345}, \,A_{1\,70}= {\cal A}^{2345 6789}, \ldots$; 
$x_1= x_{2345}, \ldots , x_{70}= x_{6789}$; and 
$H_{12345}=n_1\,,\ldots , H_{16789}=n_{70}$.
The description of $Z_{\rm zero\, modes}$ by the Riemann
theta function 
$\Theta \left[{\vec 0\atop {\vec 0}}\right] (\vec 0 , {\cal T}_{IJ})$,
defined as in [\Mumford],
with the 126x126 symmetric non-singular complex period matrix 
${\cal T}_{IJ}$ is discussed in section 3. 

To check the modular invariance of the partition function, we argue as
follows. 
In analogy with the modular group $SL(2,\cal Z)$ on the two-torus which 
can be generated by the two transformations  $\tau \rightarrow -\tau^{-1}$ 
and $\tau \rightarrow \tau + 1$,  the mapping class group of the $n$-torus,
{\it i.e.} the modular group $SL(n,\cal Z)$,
can be also generated by just two transformation as well [\Coxeter].
The  transformation
$$R_1\rightarrow R_1|\tau|,\, R_{10} \rightarrow R_{10}|\tau|^{-1},\,
\alpha \rightarrow -{|\tau|}^{-2}\alpha, \,\beta^i\rightarrow \gamma^i,\,
\gamma^i \rightarrow -\beta^i ,\, g_{ij}\rightarrow g_{ij}\eqn\modular$$
is an  $SL(10,\cal{Z})$  transformation  which leaves invariant the line
element \B\ if $d\theta^1\nobreak\rightarrow d\theta^{10}, \,
d\theta^{10} \rightarrow -d\theta^1,\, d\theta^i \rightarrow d\theta^i$.
It is the generalization of the usual $SL(2,{\cal Z})$
modular transformation $\tau \rightarrow -\tau^{-1}$.
It can be checked along the lines of [\DN] that 
\modular\ differs from a generator of $SL(10, \cal Z)$ only 
up to an $SL(9,\cal Z)$
transformation. Since we have $SL(9,\cal Z)$ invariance,
invariance under \modular\ is therefore sufficient to prove invariance under 
that $SL(10,Z)$ generator.

In order to study the properties of the partition function under \modular,
it is convenient to use the Poisson summation formula [\Green]
$$\sum_{n \in {\cal Z}^p} 
e^{-\pi (n+x)\cdot A\cdot (n+x))} = (\det A)^{-\half}
\sum_{n \in {\cal Z}^p} e^{-\pi n\cdot A^{-1}\cdot n} e^{2\pi in\cdot x}\,.
\eqn\Poisson $$ to reexpress the sum over the zero modes $n_1,...,n_{70}$.
After such resummation, we  obtain
$$\eqalign{ Z=& Z_{osc} \cdot
\sum_{n_{71},\ldots , n_{126}\in {\cal Z}^{56}}
\exp\{-\pi { R_{10}R_1 \over 5!}{\sqrt  g}g^{ii'}g^{jj'}g^{kk'}g^{pp'}g^{qq'}
H_{ijkpq}H_{i'j'k'p'q'}\cr 
& \hskip90pt -{\pi\over (4!)^2 {|\tau |}^2}{R_{10} \over
R_1}{\sqrt g} 4! g^{[jj'}g^{kk'}g^{pp'}g^{qq']} H_{ijkpq}H_{i'j'k'p'q'}
\gamma^i \gamma^{i'}\}\cr & \hskip30pt
\cdot\sum_{n_1,\ldots , n_{70}\in {\cal Z}^{70}}
\exp \{-{\pi\over (4!)^2
{|\tau |}^2} ({R_{10}\over R_1}{\sqrt g} 4! g_{[jj'}g_{kk'}g_{pp'}g_{qq']}
- i{\alpha \over g} \epsilon_{jkpqj'k'p'q'})
H_1^{jkpq}H_1^{j'k'p'q'}\cr
&\hskip70pt+ {2\pi\over 4!} ( iH_1^{jkpq}(\beta^i H_{ijkpq} + {{\cal
A}^{-1}}_{jkj'k'}\epsilon^{j'k'gh}H_{igh}\gamma^i {i\over (4!)^2} )\}
\cdot {1 \over {\sqrt{\det A}}} \cr}\eqn\almost $$  
where the $H_1^{jkpq}$ are defined to be the integers $H_{1jkpq}$,
and ${\cal A}^{-1}$ is given by
$${{\cal A}^{-1}}_{jkpqj'k'p'q'} =  |\tau|^{-2}
\bigl \{{R_{10}\over R_1}{1\over \sqrt g} \, 4! 
\,g_{[jj'}g_{kk'}g_{pp'}g_{qq']}
- i{\alpha \over g} \epsilon_{jkpqj'k'p'q'}\bigr \}
\, ,\eqn\inverse$$ 
and  $A$ defined below \zeroone\ can be proved to have $\det A=|\tau|^{70}$.
With this rewriting  \almost\ ,
it is more convenient to show that under \modular\ 
$Z_{\rm zero\,modes}$ changes according to
$$Z_{\rm zero \, modes}(R_1|\tau|, R_{10}|\tau|^{-1}, g_{ij}, 
-\alpha|\tau|^2, \gamma^i, -\beta^i)
= (\det A )^{{\scriptstyle{1\over 2}}}\,
{Z}_{\rm zero\, modes}(R_1, R_{10}, g_{ij}, \alpha, \beta^i, \gamma^i)\,.
\eqn\si$$
Therefore \modular\ generates an anomaly that 
needs to be cancelled by the sum over the oscillators.
(The determinant $\det A = |\tau|^{70}$ does not depend on the other 
parameters of the torus metric since $A$ and $A^{-1}$ are effectively
proportional to each other,  
such that  $1= \det A \, \det {A^{-1}} = (\det A )^2 |\tau|^{-140}$.)

The other $SL(10, {\cal Z)}$ generator, the analogue of
$\tau \rightarrow \tau + 1 $, instead  leaves invariant both the zero modes 
and the oscillators as in [\DN], and we will not discuss it further.
\vskip20pt

\leftline{\it The Anomaly Cancellation}
To sum over the oscillators, the key point to realize is that the chiral 
four-form in ten dimensions has 35 independent degrees of freedom.
Hence  the partition function can be written as 
$$Z = Z_{\rm zero\, modes}\,\cdot\tr e^{-2i\pi \sum_{\vec p \ne 0} p_{10}
{\cal B}_{\vec p}^{\kappa\dagger} B_{\vec p}^\kappa
-\pi R_{10}\sum_{\vec p } \sqrt{G_9^{mn} p_m p_n}\,\,
\delta^{\kappa\kappa}}\eqn\AGAIN $$
where $1\le\kappa \le 35$,  and
${\cal B}_{\vec p}^{\kappa\dagger}, B_{\vec p}^\kappa $ are the 
creation and annihilation operators associated with each degree of freedom, 
obeying the canonical commutation relations
$$[{\cal B}_{\vec p}^{\kappa\dagger}, B_{\vec p'}^\lambda] =
\delta^{\kappa\lambda}\,\delta_{\vec p,\vec p'}\,.\eqn\AOSC$$
The momenta $p_l = n_l\in {\cal Z}^9$ are integers since we are compactifying
to the torus, and the momentum $p_{10}$ can be 
written explicitly in terms of the other components 
by using the equation of motion as in [\DN]
$$\eqalign{ p_{10} =& -{G^{10\,m}\over G^{10\,10}}p_m -i {\sqrt{{G_9^{mn}\over
G^{10\, 10}}p_m p_n}}
\cr = &  -\alpha p_1 - (\alpha\beta^i + \gamma^i) p_i
-i R_{10}{\sqrt{G_9^{mn}p_m p_n}} \cr}\eqn\OMEGA$$
where $2\le i\le 9\,; 1\le m,n\le 9$.

The standard Fock space computation 
$tr\omega^{\sum_p p a^\dagger_p a_p} 
=\prod_p\sum_{k=o}^\infty \langle k |\omega^{p a^\dagger_p a_p} | k\rangle
=\prod_p {\textstyle 1\over {1 - \omega^p}}$
is used to do the trace on the oscillators in \AGAIN\ giving
$$\eqalign{Z =& Z_{\rm zero\,modes}\cdot \bigl ( e^{-\pi R_{10} \sum_{\vec n} 
\sqrt{G_9^{lm} n_l n_m}}\, \prod_{\vec n\ne \vec 0}
{\textstyle 1\over{1- e^{-i2\pi p_{10}}}}\bigr )^{35}\,.\cr}\eqn\OPF$$
\OPF\ is manifestly $SL(9,{\cal Z})$ invariant since $p_{10}$ is.
The vacuum energy $\sum_{\vec n}\sqrt{G_9^{lm} n_l n_m}$
is a divergent sum which we regularize in a way that
preserves the $SL(9,{\cal Z})$ invariance, similar to the calculation 
performed in [\DN].

The regularized version is then 
$$\eqalign{Z =& Z_{\rm zero\,modes}\cdot
\bigl ( e^{ - R_{10} \pi^{-5} 12 \sum_{\vec n\ne \vec 0} {\sqrt{G_9}\over
(G_{lm}n^ln^m)^5}}\,
\prod_{\vec n\ne \vec 0}
{\textstyle 1\over{1- e^{-2\pi R_{10}\sqrt{G_9^{lm}n_l n_m} + i 2\pi\alpha n_1
+ i 2\pi (\alpha\beta^i +\gamma^i)n_i}}}\bigr )^{35}\cr}\,.\eqn\SLFIVE$$
\noindent $Z_{\rm zero\, modes}$ is given in \zeroone\  .
We separate the product on
$\vec n = (n,n_{\perp})\ne \vec 0$ into a product
on (all $n$, $n_{\perp} \ne 0$)  and on ($n\ne 0, n_{\perp} =0$),
where $n_{\perp} = n_i$ is the momentum on the eight torus. 
Then \SLFIVE\ becomes  $$\eqalign{Z =& Z_{\rm zero\,modes}
\cdot \bigl ( e^{\textstyle{R_{10}\over\pi R_1}\zeta (2)} 
\prod_{n_1\ne 0}
{\textstyle 1\over{1- e^{2\pi i (\alpha n_1 + i {\textstyle {R_{10}\over R_1}} 
|n_1|)}}}\bigr )^{35}
\cr &\cdot
\bigl ( \prod_{n_i\ne (0,0,0,0,0,0,0,0)} 
 e^{-2\pi R_{10} <H>_{n_\perp}}
\, \prod_{n_1\in {\cal Z}}
{\textstyle 1\over{1- 
e^{-2\pi R_{10} \sqrt{G_9^{lm}n_l n_m} + i 2\pi\alpha n_1
+ i 2\pi (\alpha\beta^i +\gamma^i)n_i}}}\bigr )^{35}\cr
=& Z_{\rm zero\,modes}\cdot 
\bigl (\eta (\tau) \bar\eta(\bar\tau)\bigr )^{-35}\cr
&\cdot \bigl ( \prod_{n_i\ne (0,0,0,0,0,0,0,0)} 
e^{-2\pi R_{10} <H>_{n_\perp}}
\prod_{n_1\in {\cal Z}}
{\textstyle 1\over{1- e^{-2\pi R_{10} 
\sqrt{G_9^{lm}n_l n_m} + i 2\pi\alpha n_1
+ i 2\pi (\alpha\beta^i +\gamma^i)n_i}}}\bigr )^{35}\cr}\eqn\SLTWO$$
\noindent where $\tau\equiv \alpha + i{\textstyle {R_{10}\over R_1}}$
and 
$$<H>_{n_\perp} = - |n_\perp |^2 R_1 \sum_{n^1=1}^\infty
{\rm cos}(n_\perp\cdot\beta 2\pi n^1)
[ K_2(2\pi n^1 R_1 |n_\perp |) - K_0(2\pi n^1 R_1 |n_\perp |) ] \eqn\BES $$ 
as given in appendix A of [\DN], 
but with $n_\perp$ now given by the eight-vector
$n_i$.
\noindent In \SLTWO\ we have separated the contribution of the `2d
massless' scalars (with  zero momentum $n_{\perp}=0$) from the contribution of the
`2d massive' scalars. The latter modes are
associated with the eight-torus momentum $n_{\perp} \ne 0$ and correspond to
massive bosons on the 2-torus. Their partition function at fixed $n_{\perp}$ 
is $$e^{-2\pi R_{10} <H>_{n_\perp}} 
\prod_{n_1\in {\cal Z}}
{\textstyle 1\over{1- e^{-2\pi R_{10} \sqrt{G_9^{lm}n_l n_m} 
+ i 2\pi\alpha n_1
+ i 2\pi (\alpha\beta^i +\gamma^i)n_i}}}\eqn\MB$$
and is $SL(2,{\cal Z})$ symmetric by itself, since there is no anomaly
for massive states. The only piece of \SLTWO\ that has an
$SL(2,{\cal Z})$ anomaly is the one associated with the `2d
massless' modes 
$$e^{\textstyle{R_{10}\over\pi R_1}\zeta (2)}
\prod_{n_1\ne 0}
{\textstyle 1\over{1- e^{2\pi i (\alpha n_1 + i {\textstyle {R_{10}\over R_1}}
|n_1|)}}} = \bigl(\eta (\tau) \bar\eta(\bar\tau)\bigr )^{-1}$$
where the Dedekind eta function 
$\eta (\tau) \equiv 
e^{\pi i\tau\over{12}}\prod_{n=1}^\infty (1 - e^{2\pi i \tau n)}$,
and  $\zeta(2)={\textstyle{\pi^2\over 6}}$ 
comes from regularizing the divergent sum $\sum_n |n|$. 
Under the $SL(2,{\cal Z})$ transformation \modular\ of sect. 3,
$\tau\rightarrow -{1\over \tau}$ and  
$$(\eta (\tau) \bar\eta(\bar\tau))^{-35}\rightarrow 
|\tau|^{-35} (\eta (\tau) \bar\eta(\bar\tau))^{-35}\,. \eqn\cinque$$

This is how the oscillator anomaly cancels the zero mode anomaly
in \si. Hence the combination $Z_{\rm zero\, modes} \cdot
(\eta (\tau) \bar\eta(\bar\tau))^{-35}$ is $SL(2,{\cal Z})$ invariant,
and the expression for the partition function $Z$ derived in 
\SLTWO\ is  $SL(10,{\cal Z})$ invariant.

\chapter {$SL(2p, {\cal Z})$ Automorphic Form}
 
Following the generalization of the last section, we can now
give the partition function for the self-dual p-form on $T^{2p}$ 
for odd $p>1$. It is invariant under $SL(2p, {\cal Z})$. 
The number of physical degrees of
freedom $F$ of the chiral $(p-1)$-form abelian gauge field
with self-dual field strength $H_{i_1\ldots i_p}$ compactified on
$T^{2p}$ for odd $p>1$ is  $F = \half{(2p-2)! \over {((p-1)!)^2}}$.
The total number of zero modes is $P  = \half {(2p)! \over {(p!)^2}}$,
which is half the middle Betti number on the torus,
$b_p(T^{2p}) =\dim H^p(T^{2p}; {\cal Z}) = ({2p\atop p}) = 2P$.
Viewing $T^{2p}$ as $T^2\times T^{4k}$, we see the oscillator trace 
again breaks up into a product over 
`2d-massless scalars' given by $(\eta(\tau) \bar\eta(\bar \tau))^{-F}$
and `2d-massive' scalars so that 
$$\eqalign{Z&(R_1, R_{2p}, g_{ij}, \alpha,\beta^i, \gamma^i)
=tr  \exp\{-t{\cal H} + i2\pi\alpha P_1 + i2\pi (\alpha\beta^i +
\gamma^i)P_i\}\cr
=&Z_{\rm zero\,modes}\cdot
\bigl (\eta (\tau) \bar\eta(\bar\tau)\bigr )^{-F}\cr
&\cdot \bigl ( \prod_{n_i\ne (0,\ldots ,0)}
e^{-2\pi R_{2p} <H>_{n_\perp}}
\prod_{n_1\in {\cal Z}}
{\textstyle 1\over{1- e^{-2\pi R_{2p} 
\sqrt{G_{2p-1}^{lm}n_l n_m} + i 2\pi\alpha n_1
+ i 2\pi 
(\alpha\beta^i +\gamma^i)n_i}}}\bigr )^{F}\cr}\eqn\SLTWOp$$
where
$$\eqalign{Z_{\rm zero\, modes}  =
\sum_{n_{2F+1},\ldots , n_{P}\in {\cal Z}^{P-2F}}&
\exp\{-\pi { R_{2p}R_1 \over p!}{\sqrt  g}g^{i_1i'_1}\ldots g^{i_p i'_p}
H_{i_1\ldots i_p}H_{i_1'\ldots i'_p}\cr
& \hskip20pt -{\pi\over (p-1)! {|\tau |}^2}{R_{2p} \over   
R_1}{\sqrt g} \, g^{[j_1 j'_1}\ldots g^{{j_{p-1} j'_{p-1}}]}
H_{ij_1\ldots j_{p-1} }H_{i'j'_1 \ldots j'_{p-1}}
\gamma^i \gamma^{i'}\}\cr
&\cdot\sum_{n_1,\ldots , n_{2F}\in {\cal Z}^{2F}}
e^{-\pi (n+x)\cdot A\cdot (n+x)} \cr}\eqn\benep$$
with
$$\eqalign{{\cal A}^{j_1\ldots j_{p-1}j'_1\ldots j'_{p-1}}
= {R_{2p}\over R_1}&{\sqrt g}\,
(p-1)!\,  g^{[j_1 j'_1}\ldots g^{{j_{p-1} j'_{p-1}}]}
+ i\alpha \epsilon^{j_1\ldots j_{p-1}j'_1\ldots j'_{p-1}}\,,\cr
x_{j_1\ldots j_{p-1}} \equiv \beta^i H_{ij_1\ldots j_{p-1}}& 
+ {i \over ((p-1)!)^2}\gamma^i
{{\cal A}^{-1}}_{j_1\ldots j_{p-1}j'_1\ldots j'_{p-1}}
\epsilon^{j'_1\ldots j'_{p-1} k_1\ldots k_{p-1}} H_{i k_1\ldots k_{p-1}}\,,\cr
{{\cal A}^{-1}}_{j_1\ldots j_{p-1}j'_1\ldots j'_{p-1}} =  |\tau|^{-2}
&\bigl \{{R_{2p}\over R_1}{1\over \sqrt g} \, (p-1)! \,g_{[j_1j'_1}\ldots
g_{j_{p-1}j'_{p-1}]}
- i{\alpha \over g} \epsilon_{j_1\ldots j_{p-1}j'_1\ldots j'_{p-1}}
\bigr \}\,\cr}$$    
and the transverse directions corresponding to directions on $T^{4k}$
are $i_1, j_1,\ldots = 2, \ldots ,2p-1$, for $p=1+2k$.   

Our proof shows the invariance of \SLTWOp\
under the generators of $SL(2p, {\cal Z})$ which are given by the 
two transformations generalizing $\tau\rightarrow \tau -1$
and $\tau\rightarrow -{1\over\tau}$:
$$\eqalign{R_1&\rightarrow R_1 , R_{2p}  \rightarrow R_{2p}, \alpha
\rightarrow \alpha - 1, \beta^i\rightarrow \beta^i,
\gamma^i \rightarrow \gamma^i+\beta^i, g_{ij}\rightarrow
g_{ij}\,;\cr
R_1&\rightarrow R_1|\tau|,\, R_{2p}  \rightarrow R_{2p}|\tau|^{-1},\, \alpha
\rightarrow -{|\tau|}^{-2}\alpha, \,\beta^i\rightarrow \gamma^i,\,
\gamma^i \rightarrow -\beta^i ,\, 
g_{ij}\rightarrow g_{ij}\,. \cr}\eqn\modular$$
We can express \SLTWOp\ as
$Z(R_1, R_{2p}, g_{ij}, \alpha,\beta^i, \gamma^i)
= {\Theta \over \Delta}$
where our explicit calculation of $\Delta$ gives
$$\eqalign{{\hskip-30pt}\Delta & =
(\eta (z)\bar\eta (\bar z))^{{1\over 2}{(2p-2)!\over {((p-1)!)^2}}}\cr
&\cdot\,( \prod_{n_i\ne (0,\ldots ,0)}
e^{-2\pi R_{2p} <H>_{n_\perp}}
\prod_{n_1\in {\cal Z}}
{\textstyle 1\over{1- e^{-2\pi R_{2p} 
\sqrt{G_{2p-1}^{lm}n_l n_m} + i 2\pi\alpha n_1
+ i 2\pi
(\alpha\beta^i +\gamma^i)n_i}}}\bigr )^{-{1\over 2}{(2p-2)!\over 
{((p-1)!)^2}}\,\,}\,.\cr}\eqn\massive$$
$\Delta$ is the trace over the non-zero modes of the
self-dual $G_p$ on the manifold $T^{2p}$, for odd $p>1$. 
For a different manifold $M^{2p}$, the
function $\Delta$ would be different.     

The zero mode contribution  $Z_{\rm zero\, modes}$ is a 
Riemann theta function (with zero-valued characteristics)
$\Theta \left[{\vec 0\atop {\vec 0}}\right] (\vec 0 , {\cal T})
= \sum_{n_1,\ldots , n_P\in {\cal Z}^P}
\exp\{i \pi { n^I n^J {\cal T}_{IJ}}\}$
where the rank $P$ complex symmetric period matrix ${\cal T}_{IJ}$
can be reconstructed from \benep . 
This particular theta function is one of the  $2^{2P}$ spin structures on
a Riemann surface of genus P. (A Riemann surface $\Sigma_g$ of genus $g$ has
$2^{2g}$ spin structures which transform 
into each other under $Sp(2g,{\cal Z})$,
the mapping class group of $\Sigma_g$). 
According to our proof, this particular theta
function $\Theta \left[{\vec 0\atop {\vec 0}}\right] (\vec 0 , {\cal T})$
multiplied by $(\eta(\tau) \bar\eta(\bar \tau))^{-F}$
is actually invariant under an $SL(2p,{\cal Z})$ subgroup of
$Sp(2P,{\cal Z})$. (On $T^n$, there are $2^n$ spin structures, which
in general transform into each other under $SL(n,{\cal Z})$).

\chapter {Calculation of the Period Matrix from A- and B-cycles}
We now discuss our result in terms of the description via 
the period lattice and omega function given in 
[\Wittentwo,\WM ]
for calculating partition functions on arbitrary spacetimes of dimension 
$2+4k$. 
We identify the period matrix in our expression, and show how
for the simple case of a rectangular torus, it indeed could have been
constructed directly from the A- and B-cycle periods. 
For the general flat metric, however, our formula 
provides a simple expression for the B-lattice basis vectors (on the torus),
which otherwise would be more complicated to compute directly. 

To this end,
we focus on the theta function given by our 
formula in the simplest case of the
fivebrane chiral two-form ($p=3$). In general,  
our $p$-forms are closed, since they are locally exact $G_p =dB$, and
their periods ${1\over {2\pi}} \int_{U_p} G_p$,
defined with respect to $p$-cycles $U_p$ 
will take values in the period lattice
$\Lambda = H^p(T^{2p};{\bf Z}) = {\bf Z}^{b_p(T^{2p})} =
{\bf Z}^{({2p\atop p})} = {\bf Z}^{2P}$,
where $2P \equiv {({2p\atop p})}$.
For the fivebrane on $T^6$, the period lattice is 
$\Lambda = H^3(T^6;{\bf Z}) = {\bf Z}^{20}$.
The lattice $\Lambda = {\bf Z}^{20}$ has $2^{20}$ theta functions  
and the number of zero modes is $P = 10$.  The $2^{20}$ theta
functions correspond to those defined on the Riemann surface $\Sigma_{10}$
of genus $g = 10$. $Z_{\rm zero\,modes}$ is given by the one that has 
zero characteristics and is invariant (up to a c-number multiple) under
$SL(6,{\cal Z})$. 
To display its $10\times 10$ period matrix ${\cal T}_{IJ}$ we find, from 
rewriting  [\DN] the explicit sum over the zero modes in \SLTWOp\ , that
the relevant theta function is $$\eqalign{Z_{\rm zero\, modes}  
&=\sum_{n_1,\ldots , n_{10}\in {\cal Z}^{10}}
\exp\{-\pi { R_6R_1 \over 6}{\sqrt  g}g^{ii'}g^{jj'}g^{kk'}
H_{ijk}H_{i'j'k'}\cr
&\hskip80pt -{\pi\over 4} [{R_6\over R_1}{\sqrt g}(g^{jj'}g^{kk'}
- g^{jk'}g^{kj'}) + i\alpha \epsilon^{jkj'k'}]H_{1jk}H_{1j'k'}\cr
&\hskip80pt -{\pi\over 2}[{R_6\over R_1}{\sqrt g}(g^{jj'}g^{kk'} -
g^{jk'}g^{kj'}) + i\alpha \epsilon^{jkj'k'}]
\beta^i H_{1jk}H_{ij'k'}\cr
&\hskip80pt -{\pi\over 4}[{R_6\over R_1}{\sqrt g}(g^{jj'}g^{kk'} -
g^{jk'}g^{kj'}) + i\alpha \epsilon^{jkj'k'}] \beta^i
\beta^{i'}H_{ijk}H_{i'j'k'}\cr
&\hskip80pt
-{i\pi \over 2} \gamma^i \epsilon^{jkj'k'} H_{1jk} H_{ij'k'}\}\cr
=&\sum_{n_1,\ldots , n_{10}\in {\cal Z}^{10}}
\exp\{i \pi { n^I n^J {\cal T}_{IJ}}\}           
\,= \,\sum_{x\in \Lambda_1} \Omega (x) e^{i\pi (x,{\cal T} x)}\,
= \,\Theta (\vec 0, {\cal T}_{IJ}) \cr}\eqn\period$$

\noindent where $H_{123}=n_1$,
$H_{124}=n_2$, $H_{125}=n_3$, $H_{134}=n_4$, $H_{135}=n_5$,
$H_{145}=n_6$, $H_{234}=n_7$, $H_{235}=n_8$, $H_{245}=n_9$,
$H_{345}=n_{10}$. So we identify $\Lambda_1 = {\bf Z}^{10}$,
the omega function restricted to $\Lambda_1$ as
$\Omega(x) =1$, and the $10\times 10$
period matrix ${\cal T}_{IJ}$ in terms of the parameters of the
metric, {\it i.e.} ${\cal T}_{11} =
i {R_6\over R_1} {\sqrt g} (g^{22}g^{33} - g^{23}g^{32})$, etc.
For an arbitary flat metric on $T^6$, all components of 
${\cal T}_{IJ}$ are non-vanishing. 

For simplicity, we reduce to the case of a metric for a rectangular torus.
Then $ G_{11} = R_1^2$, $g_{22} = R_2^2$,
$g_{33} = R_3^2$, $g_{44} = R_4^2$, $g_{55} = R_5^2$, $G_{66} = R_6^2$ (and
all other parameters $g_{ij}=0$ for $i\ne j$, and $\alpha, \beta^i,
\gamma^i = 0$. ) The period matrix which occurs in \period\ then
has only non-vanishing diagonal components given by
$$\eqalign{{\cal T}_{11} &= i {{R_6 R_5 R_4}\over {R_1 R_2 R_3}},\qquad
{\cal T}_{22} = i {{R_6 R_3 R_5}\over {R_1 R_2 R_4}},\cr
{\cal T}_{33} &= i {{R_6 R_3 R_4}\over {R_1 R_2 R_5}},\qquad
{\cal T}_{44} = i {{R_6 R_2 R_5}\over {R_1 R_3 R_4}},\cr
{\cal T}_{55} &= i {{R_6 R_2 R_4}\over {R_1 R_3 R_5}},\qquad 
{\cal T}_{66} = i {{R_6 R_2 R_3}\over {R_1 R_4 R_5}}\cr
{\cal T}_{77} &= i {{R_1 R_6 R_5}\over {R_2 R_3 R_4}},\qquad 
{\cal T}_{88} = i {{R_1 R_6 R_4}\over {R_2 R_3 R_5}}\cr     
{\cal T}_{99} &= i {{R_1 R_6 R_3}\over {R_2 R_4 R_5}},\qquad 
{\cal T}_{10\,10} = 
i {{R_1 R_6 R_2}\over {R_3 R_4 R_5}}\,.\cr}\eqn\perioddiag$$

Alternatively, we can  
derive these components of the period 
matrix directly by picking the `B-periods' as follows.
Using the method of [\Wittentwo], we construct the period matrix
of the theta function by
noting that for self-dual quantum fields we should sum over only
half the periods, {\it i.e.} those which lie on the A-lattice. 
In this case, from  [\Wittentwo] we find the lattice of 
A periods is $\Lambda_1 = H^3(T^5; {\bf Z}) = {\bf Z}^{10}$, and 
the lattice of B periods is $\Lambda_2 = H^2(T^5; {\bf Z}) = {\bf Z}^{10}$,
where $\Lambda =  \Lambda_1 \oplus \Lambda_2$.
But the values the B periods take on reflect the choice of basis vectors
of the B-lattice, {\it i.e.} the periods of the B-cycles lie on the lattice
$\Lambda_B = \sum_{I=1}^{10} n_I {\cal T}_{IJ}$ where $ n_I \in Z^{10}$.
That is to say, the values of the periods
$\int_{U_3} H_{LMN} dx^L \wedge dx^M \wedge dx^N$  lie in  $\Lambda_B$,
when $U_3$ is a B-cycle, for eg. one of ten possible basic cycles 
corresponding to $H_{6MN}$. 

The A-cycles (whose periods lie on the integer lattice 
$\Lambda_1 = \sum_{I=1}^{10} n_I \delta_{IJ}$)
correspond to the other ten components of $H_{LMN}$, i.e. $H_{321}, 
H_{421}, \dots H_{543}$.    

{}From the self-dual equation of motion [\DN], we  have for instance 
$H_{654} 
= {1\over\sqrt{-G}} G_{66} G_{55} G_{44} \epsilon^{654321} \, H_{321}$,
i.e. $$H_{654} = i  {{R_6 R_5 R_4}\over {R_1 R_2 R_3}}   H_{321}\,.\eqn\sd$$

Therefore, if the period over the A-cycle correponding to $H_{321}$ is $n$,
then the period of the B-cycle corresponding to  $H_{654}$   
is ${\cal T}_{11} n$
where  ${\cal T}_{11} = i  {{R_6 R_5 R_4}\over {R_1 R_2 R_3}}$. 
This last sentence
is due to the self-dual equation 
for  $H_{654}$ given in \sd\ above. So this gives us a direct computation
of ${\cal T}_{11}$, and it matches what we found from our calculation of
the partition function \perioddiag\ . (All components ${\cal T}_{IJ}$
follow similarly ).

\chapter{Conclusions}

In this paper we computed explicitly the partition function 
$Z={\Theta\over\Delta}$ of 
the type IIB superstring chiral four-form on a ten-torus, and
extended the result 
for the general chiral $(p-1)$-form on $T^{2p}$ for all odd $p>1$. 
We found expressions for
the determinant of the non-zero modes $\Delta$, and the period matrx
${\cal T}_{IJ}$ occuring in the theta function describing the zero modes,
in terms of the metric parameters of the spacetime torus.
We also showed how this period matrix comes in from a direct calculation
via the cycles.

In analogy with the  $SL(2,{\cal Z})$ modular invariance
of 2d string theory, the partition function of the self-dual $p$-form
field strength has $SL(2p, {\cal Z})$ modular invariance on
$T^{2p}$ for odd $p>1$. Invariance under the mapping class group of
the spacetime would not necessarily be the case for more general
manifolds.
In general, a knowledge of the partition functions for self-dual p-forms
on various manifolds may lead to a better formulation of these
non-Lagrangian fields and ultimately the description of unitarity
in M-theory. 

\vskip20pt

\refout

\end